# High power GaSb-based distributed feedback laser with laterally coupled dielectric gratings at 1.95 μm

*Zhengqing Ding[1#], Juntian Cao[2,3#], Kun Zhan[1], Yihang Chen[2,3], Lidan Zhou[1],Hao Tan[4,5], Chenao Yang[2,3]*, Ying Yu[1]*, Zhichuan Niu[2,3]*, Siyuan Yu[1]*

[1] State Key Laboratory of Optoelectronic Materials and Technologies, School of Electronics and Information Technology, Sun Yat-Sen University, Guangzhou 510275, China

[2] Key Laboratory of Optoelectronic Materials and Devices, Institute of Semiconductors, Chinese Academy of Sciences, Beijing, China

[3] Center of Materials Science and Optoelectronics Engineering, University of Chinese Academy of Sciences, Beijing, China

[4] International Quantum Academy, Shenzhen, China

[5] Hefei National Laboratory, Hefei, China

**Abstract:** Traditional Distributed Feedback (DFB) or Distributed Bragg Reflector (DBR) lasers typically utilize buried gratings as frequency-selective optical feedback mechanisms. However, the fabrication of such gratings often necessitates regrowth processes, which can pose technical challenges for materials platforms such as GaAs and GaSb. Metal gratings were also used for GaSb lasers but they introduce additional absorption loss that limits device efficiency and output power. In this paper, we introduce a novel laterally coupled dielectric Bragg grating structure, which enables highly controllable, deterministic, and stable coupling between the grating and the optical mode. Our device demonstrates a continuous-wave output power of 47.02 mW at room temperature, exhibiting stable single-mode operation from 300-1000 mA and achieving a maximum side mode suppression ratio of 46.7 dB. These results underscore the innovative lateral coupled dielectric grating as a feasible and technologically superior approach for fabricating DFB and DBR lasers, which hold universal applicability across different material platforms and wavelength bands.

## Introduction

Antimony-based III-V compound materials (such as InGaSb/GaSb, InAs/GaSb etc.), with their inherent narrow bandgap, natural lattice matching, and tunable band structures, are primary candidates for infrared (2-4 μm) optoelectronic systems, with promising applications in gas detection [1], remote sensing [2] and laser spectroscopy [3]. In pursuit of high-power, high-spectral-purity single-mode laser performance, Bragg gratings are often employed to provide selective optical feedback, facilitating the development of distributed feedback (DFB) or distributed Bragg reflector (DBR) lasers. For lasers based on indium phosphide (InP) operating in 1.3-1.7 μm, gratings are typically etched on top of the active layer, followed by a regrowth process to bury the gratings and complete the epitaxial structure [4]–[6]. However, this process is challenging for GaSb system due to the oxidation of high-Al composition in the upper cladding layer.

A regrowth-free alternative is the laterally coupled distributed feedback (LC-DFB) laser, with Bragg gratings placed alongside the ridge waveguide, using either etched [7]–[10] or metal gratings [11]–[13]. Metal gratings introduce gain coupling and eliminate the degenerate modes of index-coupled gratings, improving single-mode stability and higher side mode suppression ratio (SMSR). However, they cause additional absorption loss, limiting laser output power to around 10 mW [11], [12]. Etched gratings in GaSb-based materials face a significant challenge compared to InP-based laser structures due to the lack of a suitable selective etch-stop layer, which would allow precise control over the etching depth just above the active layer [14]. Without this layer, an undesirable feature known as "footing" may occur, leading to a non-deterministic grating coupling coefficient (κ) and unstable single-mode operation.

In this paper, for the first time, we utilize dielectric gratings [15] to fabricate GaSb-based LC-DFB lasers. We first optimize the etching process to reduce the footing on both sides of the ridge waveguide to below 100 nm, enhancing interaction with the optical field. Amorphous silicon (α-Si) was subsequently used to form a high refractive index contrast grating (Δn~1.5) on both sides of the ridge waveguide, providing sufficient coupling strength for high-quality single-mode spectra. We achieved a high-power 1.95 μm dielectric grating LC-DFB laser with a maximum output power of 47.02 mW at room temperature. The maximum SMSR was 46.7 dB, and the device maintained single-mode operation over a current range of 300-1000 mA, demonstrating excellent spectral purity. Our device, which does not require a regrowth process, achieved a deterministic and stable κ, confirming the universality of dielectric gratings in material systems without etch-stop layers. This paves the way for future high-quality single mode laser sources.

## Laser Design and Fabrication

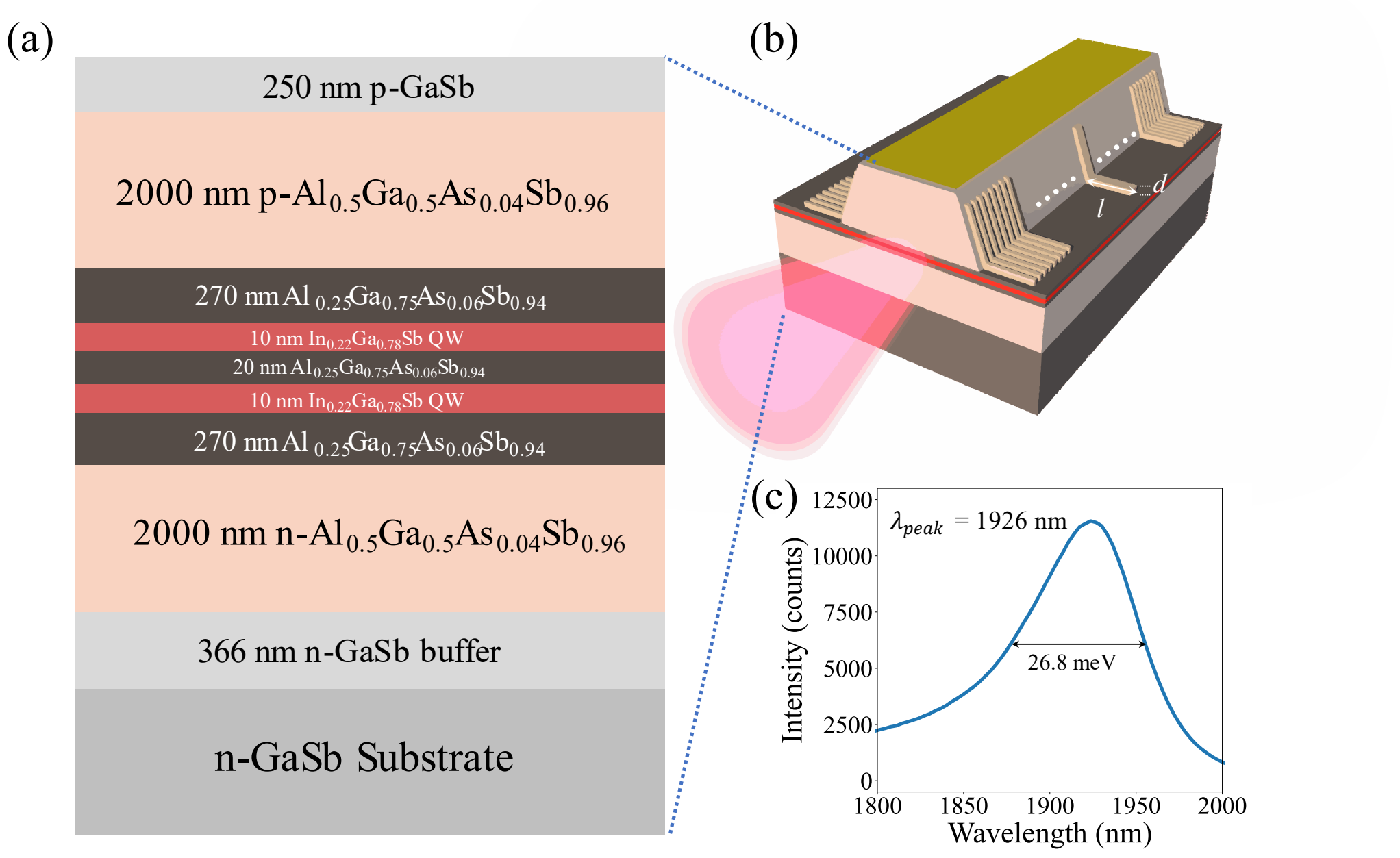

Figure 1. **Epitaxial wafer structure and LC-DFB laser schematic.** (a) A diagram illustrating the layer stack of the epitaxial wafer, which is the foundational structure for our semiconductor laser device; (b) Schematic of the laterally coupled distributed feedback (LC-DFB) laser, featuring a trapezoidal waveguide and α-Si gratings. The α-Si gratings are patterned to provide low-loss and effective mode selection, while the trapezoidal waveguide ensures efficient light confinement and zero footing; (c) The photoluminescence spectrum of the QW active layers.

The fabrication of our GaSb laser devices starts with the epitaxial growth on an N-type GaSb (100) substrate, as illustrated in Figure 1(a). The growth sequence starts with a 366 nm N-type GaSb buffer layer, followed by a 2000 nm $Al_{0.5}Ga_{0.5}As_{0.04}Sb_{0.96}$ lower N-type cladding layer, chosen for its suitable refractive index contrast and lattice compatibility. A 270 nm undoped $Al_{0.25}Ga_{0.75}As_{0.06}Sb_{0.94}$ waveguide layer follows, ensuring efficient light confinement and carrier transport. The active region, which is symmetrically centered for optimal modal gain, consists of two pairs of 10 nm thick $In_{0.22}Ga_{0.78}Sb$ quantum wells (QWs). The epitaxial process concludes with a 2000 nm P-type $Al_{0.5}Ga_{0.5}As_{0.04}Sb_{0.96}$ upper cladding layer and a 250 nm P-type GaSb contact layer, which facilitate efficient current injection. The photoluminescence (PL) peak wavelength of the structure is 1926 nm with a narrow fullwidth at half-maximum (FWHM) of 26.8 meV as shown in Figure 1(c).

The laser device's structural design, including the trapezoidal ridge waveguide and the α-Si sidewall grating, is illustrated in Figure 1(b). The lateral modes are primarily determined by the lower width of the trapezoidal ridge waveguide, with the cutoff width for the first-order transverse mode ($TE_{01}$) at 1930 nm wavelength calculated to be 2.4 μm using finite-difference methods, as shown in Figure 2(a). This width is critical for balancing modal gain while maintaining single-mode operation. The grating coupling coefficient (κ), a critical parameter for the device's performance, is evaluated using coupled mode theory [16]:

$$\kappa = \frac{n_2^2 - n_1^2}{\lambda_0 n_{eff}} \cdot \frac{\sin(\pi m \Lambda)}{m} \cdot \Gamma \tag{1}$$

where $n_1$ and $n_2$ are the refractive index of α-Si and bisbenzocyclobutene (BCB) that encapsulates the grating, respectively. $\lambda_0$ represents the Bragg wavelength, $n_{eff}$ is the effective refractive index of the ridge waveguide transmission mode, m=1 corresponds to the diffraction order of the grating, Λ is the duty cycle, and Γ is the optical confinement factor. Figures 2(b) and 2(c) illustrate the dependence of the coupling coefficient κ on the grating deposition thickness *d* and the grating extrusion length *l*, revealing an optimal choice of *d* = 220 nm and *l* = 3 μm for high process tolerance. The total coupling strength *κL* for our *L*=3 mm long device was estimated to be ~ 4.5, providing sufficient gain and coupling strength for high power and pure single-mode lasing.

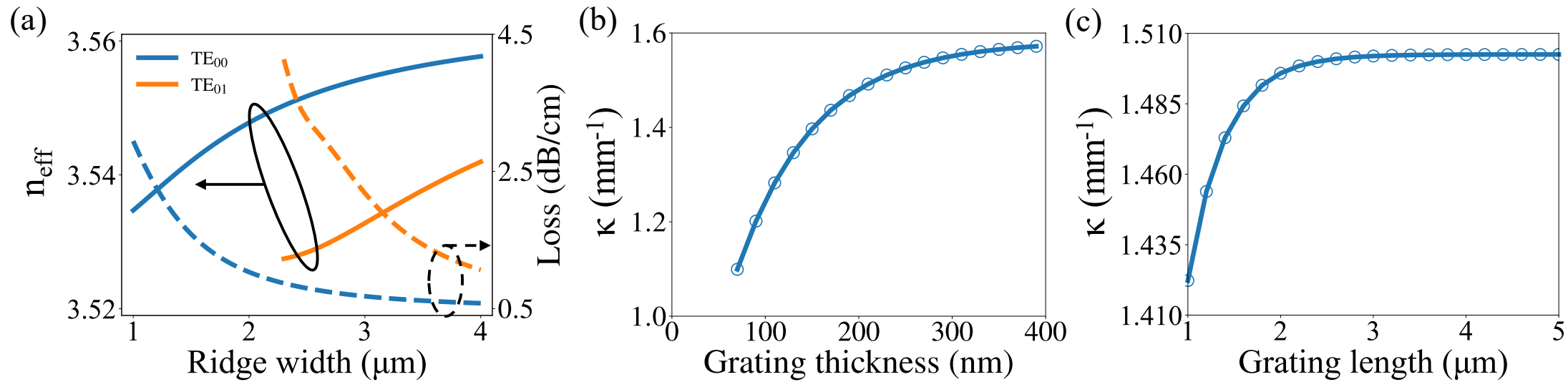

Figure 2. **Waveguide mode and gratings characteristics.** (a) The effective refractive index and waveguide loss as a function of ridge width for the $TE_{00}$ and $TE_{01}$ transverse modes; (b-c) Coupling coefficient κ as a function of grating thickness (b), and grating length (c), respectively.

Dynamic simulations using the Time-Domain Traveling-Wave (TDTW) method [17], as shown in Figures 3(a) and 3(b), demonstrate that asymmetric facet coatings can optimize the optical field distribution, potentially enhancing the device's maximum single-mode output power. The application of facet coatings enhances the output power by 1.87 times and reduces the photon density at the central λ/4 phase shift by approximately 22%, thereby improving optical field uniformity.

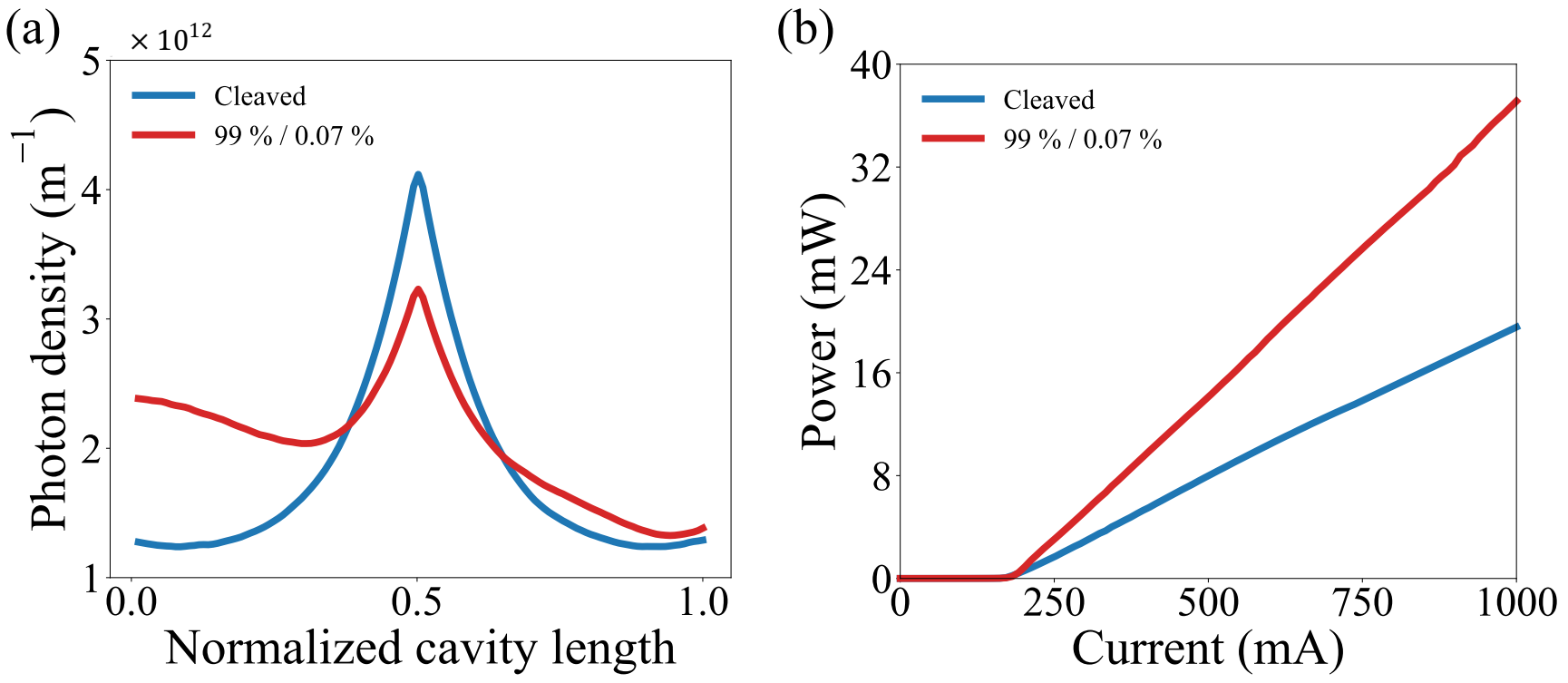

Figure 3. **Simulated DFB Laser performance under HR/AR coating.** (a) Photon density distribution within the laser at an injection current of 3.8 times the threshold current ($I_{th}$), The cleaved facet reflectivity was set at 30%; (b) The simulated single-end output power of the laser under various current injections.

In the fabrication process, the upper ridge width is precisely defined using electron beam lithography (EBL) with a hydrogen silsequioxane (HSQ) electron resist, ensuring a smooth sidewall morphology. The sidewalls are then etched using a chlorine-based ICP-RIE process, achieving a precise 79° angle with the active region plane and eliminating the footing issue. A 40 nm $SiN_x$ passivation layer and a 220-nm-thick α-Si layer are deposited by ICP-CVD and electronic beam evaporation, respectively. First-order gratings tarting the Bragg wavelength of 1940 nm, with a period of 273.26 nm and duty ratio of 60%, are patterned alongside the ridge using an EBL resist and etched with a fluoride-based RIE process. Figure 4(a) presents a SEM image of the fabricated α-Si grating a λ/4 phase shift at its center, a design critical for stabilizing single-mode operation.

Post-fabrication, the device is encapsulated with BCB for planarization. Contacts are formed with P-type Ti/Pt/Au on the contact window and N-type AuGe/Ni/Au on the backside after substrate thinning to 135 μm. As designed, the front facet is coated with a pair of $Ta_2O_5/SiO_2$ anti-reflection coatings (AR, 0.07%), and the back facet with three pairs of $SiO_2$/Si high-reflection coatings (HR, 99%) to suppress Fabry-Perot modes and enhance the DFB output power. The completed devices are then cleaved into bars and mounted onto a high thermal conductivity AlN substrate, as shown in the microscope images in Figure 4(b), to ensures good heat dissipation.

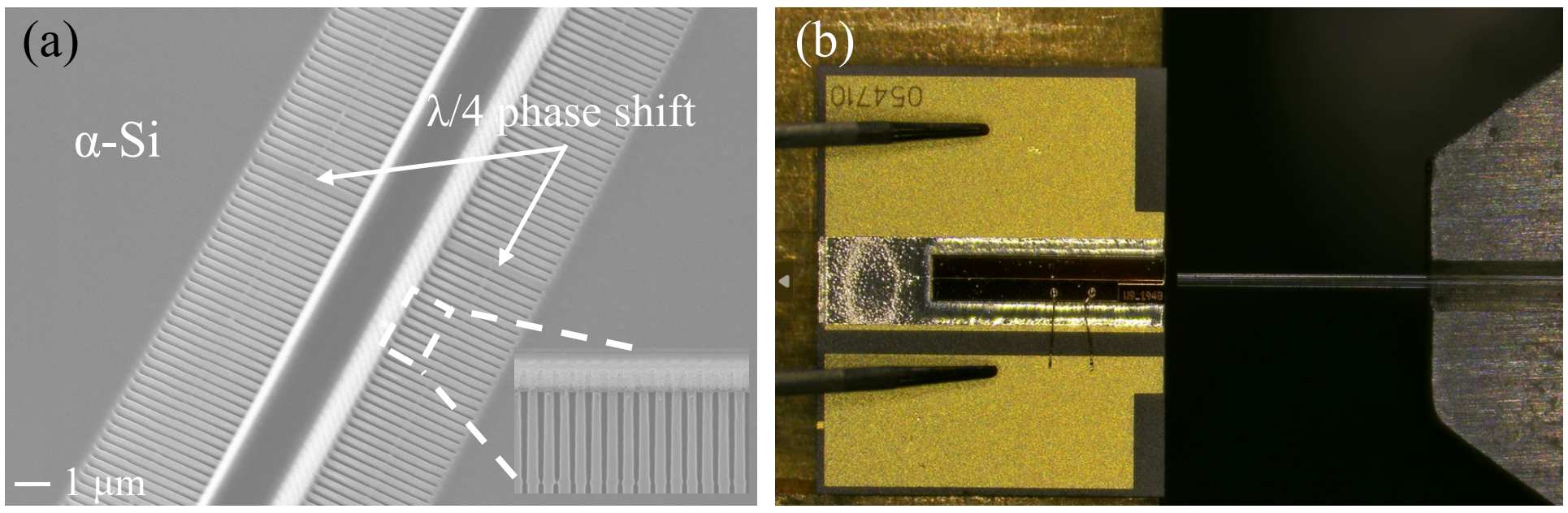

Figure 4. **Gratings fabrication and device packaging.** (a) SEM images of the etched α-Si gratings with a λ/4 phase shift in the middle, detailed information of the grating is shown in the inset; (b) Completed devices post-fabrication with CoS packaging and fiber coupling.

## Lasing Characteristics and Intrinsic Linewidth

Figure 5(a) illustrates the typical *L-I-V* characteristics of our 2.4 μm × 3 mm DFB laser at room temperature, with a maximum output power of 47.02 mW, a turn-on voltage of 0.67 V, and a differential series resistance of 0.96 Ω. The output power, measured with a Thorlabs S148C large-area power meter, exceeds previous reports[7]-[13] due to the use of dielectric grating and enhanced by the long cavity length and facet coating design. The low turn-on voltage and differential resistance significantly reduce the device's power consumption, enabling higher injection currents and less heat generation, thereby improving power conversion efficiency (PCE).

The threshold current of 262 mA corresponds to a threshold current density of $J_{th}$=3638.9 A/cm², slightly higher than previous works [9]. $J_{th}$ is decided by a number of factors including $\kappa L$, intra-cacity loss and nan-radiative recombination. In our case an over-etching of the waveguide depth due to the lack of etch end point detection (EPD) in our ICP equipment for GaSb materials could have contributed to higher non-radiative recombination losses within the QW active region, consequently increasing $J_{th}$ , which can be improved in future fabrication processes.

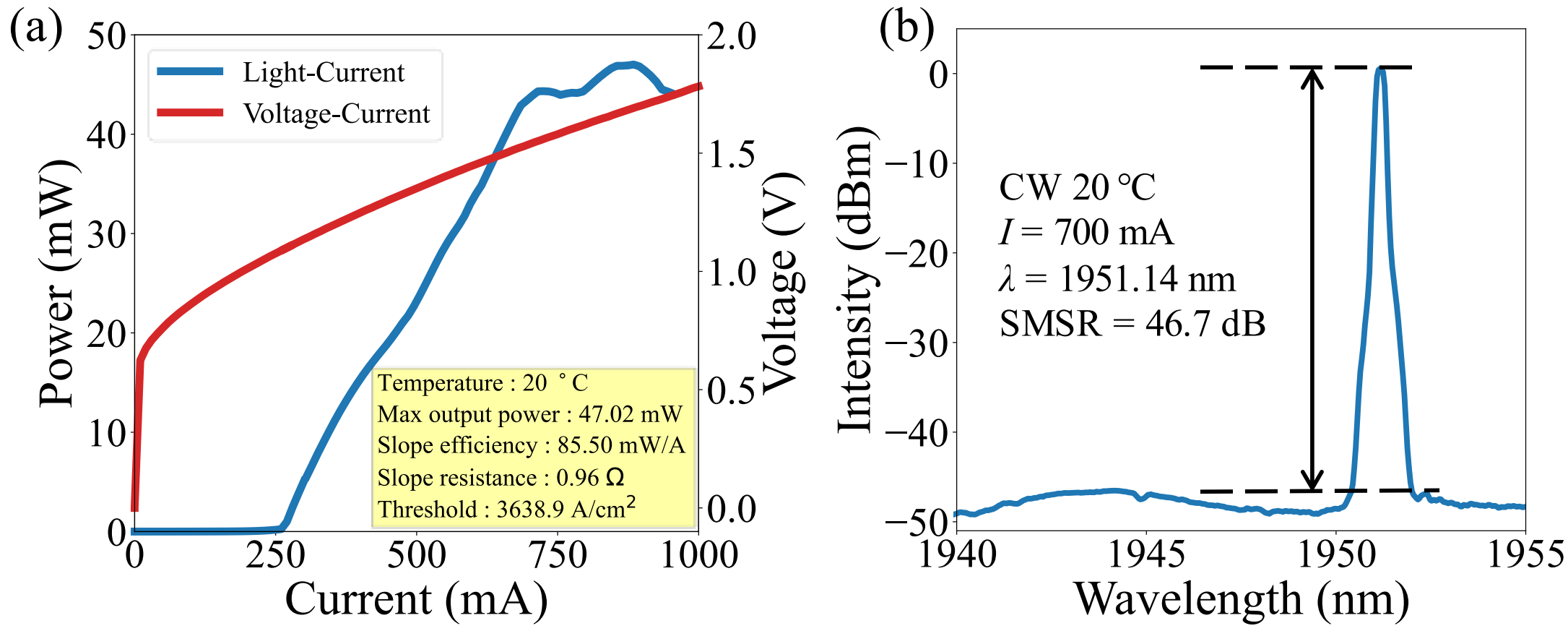

Figure.5. (a) *L-I-V* characteristics of the DFB laser at room temperature; (b) Optical spectra of the single DFB laser operating at a drive current of 700 mA.

Our device exhibits stable single-mode operation across a broad injection current range of 300-1000 mA at room temperature, as shown in Figure 6(a). The emission spectra, captured using a Yokogawa AQ6376 optical spectrum analyzer with a 1950 nm single-mode polarization-maintaining fiber, demonstrate a wavelength variation rate of 0.0094 nm/mA with current. The side mode suppression ratio (SMSR) initially increases due to the gain spectrum-Bragg wavelength mismatch but decreases with further current increase, reaching a maximum of 46.7 dB at 700 mA (2.67 × $I_{th}$), as illustrated in Figure 5(b), indicative of high spectral purity. Additionally, temperature stability from 15 to 50°C is evidenced in Figure 6(c), with the Bragg wavelength shifting from 1950 nm to 1955 nm at a rate of 0.14 nm/K with increasing thermoelectric cooler (TEC) temperature. Mode hopping near 35°C is attributed to the spatial hole burning (SHB) effect due to strong coupling ($\kappa L \approx 4.5 \gg 1$), which diminishes the threshold gain difference between modes, leading to mode competition [18].

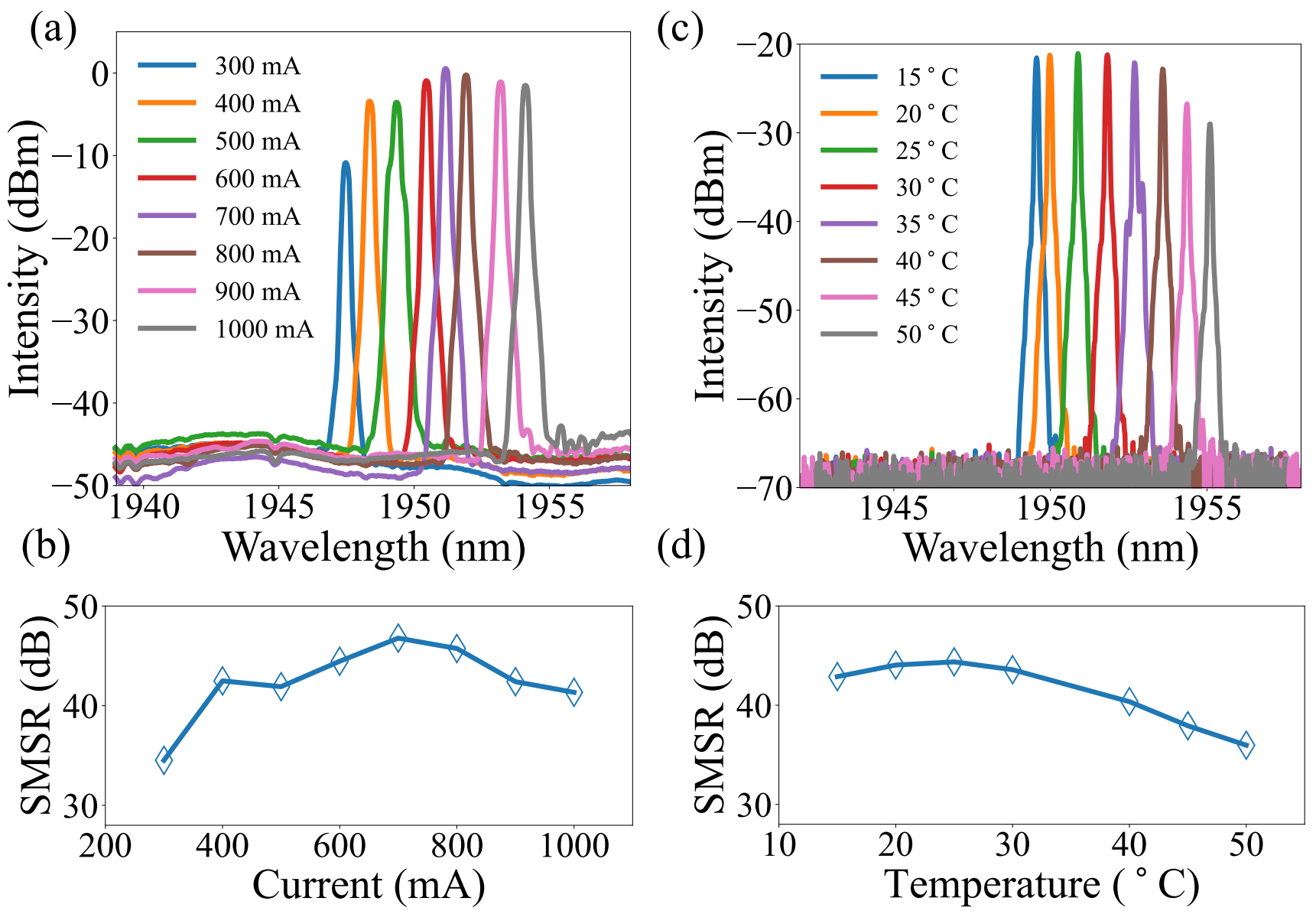

Figure.6. Emission spectra varies (a) injection current at 20℃ and (b) corresponding SMSRs; (c) temperature at 700 mA and (d) corresponding SMSRs.

The intrinsic linewidth was extracted using a commercial optical noise analysis system (OEWaves OE4000), from the high-frequency white noise beyond the 1/f noise. Figure 6 shows the frequency noise spectrum of the laser biased at 700 mA and 20°C, with the dashed line indicating the white noise level. The white noise level, calculated from the flat optical frequency noise spectral density interval between 3 MHz and 10 MHz, was $2.14 \times 10^6$ Hz²/Hz, corresponding to an intrinsic linewidth of 6.72 MHz, which is lower than in previous work.

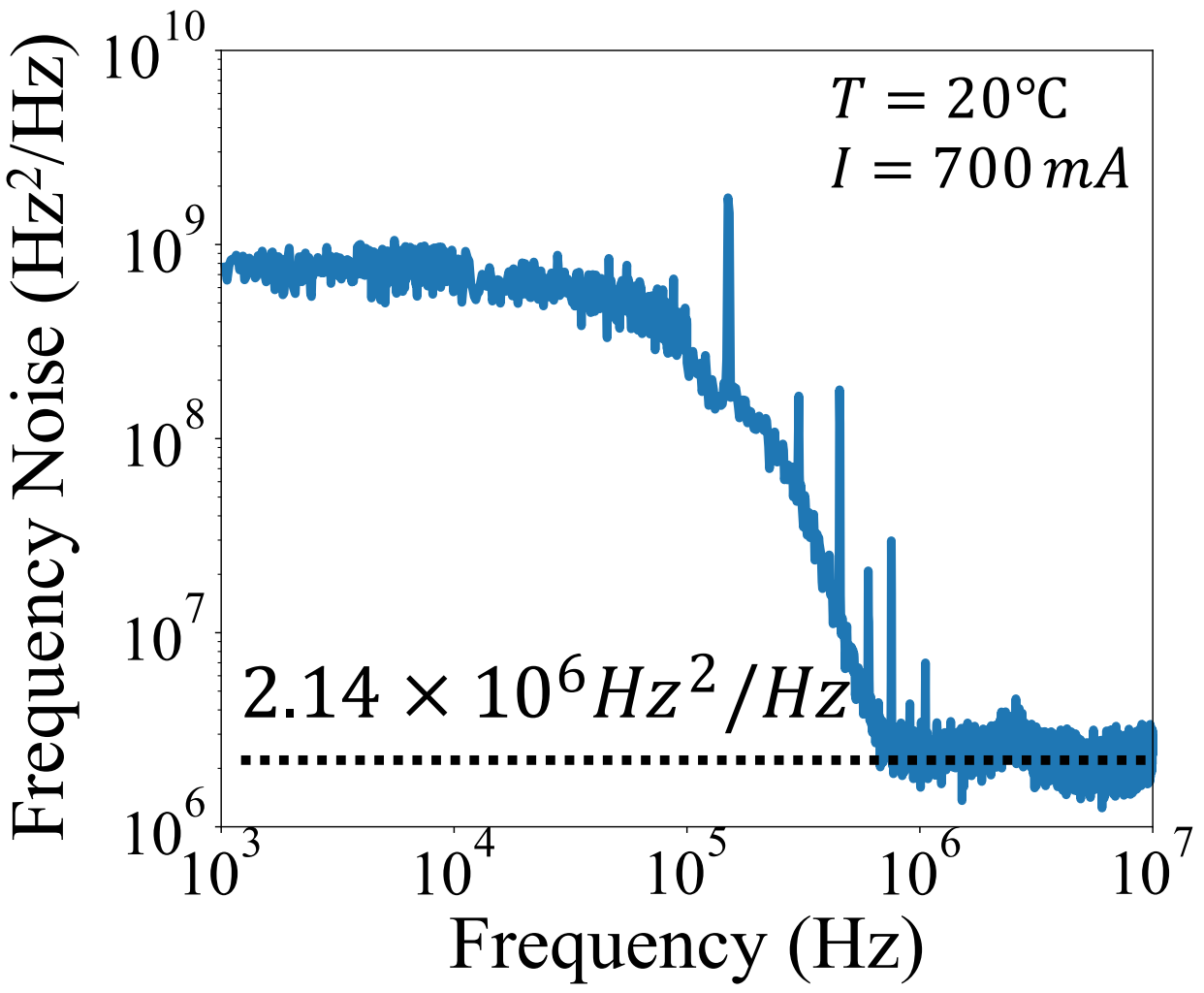

Figure 7. Frequency noise spectrum and measured white noise level at 20 ℃ and 700 mA.

Table 1. Comparison of the performance of this work with reference 2-3 μm GaSb based LC-DFB lasers.

| Year | Size (μm$^2$) | Grating material | Threshold Current Density (A/cm$^2$) | Facet coating | Power (mW) @RT | SMSR (dB) | Line-width (MHz) | Ref. |
|---|---|---|---|---|---|---|---|---|
| 1999 | 4 × 1000 | 1st order Cr | 790 | cleaved | 5@pulsed | 30 | - | [19] |
| 2004 | 4 × 800 | 1st order Cr | 938 | cleaved | 8.5 | 33 | - | [20] |
| 2006 | 4.7 × 745 | 1st order Cr | 1250 | 30%/80% | 5 | 35 | 2.2 | [11] |
| 2010 | 4 × 1200 | 1st order Cr | 255 | 30%/80% | 6 | 34 | - | [12] |
| 2010 | 1.5 × 1000 | 3rd order Cr | 6000 | cleaved | 2 | 30 | - | [21] |
| 2012 | 4 × 2000 | 2nd order GaSb | 425 | 2%/30% | 35 | 20 | 1.4 | [7] |
| 2013 | 1.5 × 2000 | 6th order GaSb | 1040 | cleaved | 40 | 25 | - | [22] |
| 2014 | 6 × 2000 | 1st order GaSb | 1500 | 5%/95% | 15 | 20 | - | [23] |
| 2016 | 3.5 × 600 | 2nd order GaSb | 2381 | 1.4%/84% | 25 | 50 | - | [24] |
| 2016 | 4 × 2000 | 2nd order GaSb | 1250 | 1%/30% | 14 | 20 | - | [8] |
| 2018 | 6 × 1500 | 21th order Cr | 1444.4 | cleaved | 14 | 13.4 | - | [9] |
| 2019 | 4.5 × 1000 | 2nd order Cr | 593 | cleaved | 40@10℃ | 53 | 9 | [13] |
| 2023 | 4 × 500 | 1st order Cr | 950 | cleaved | 1.5 | 35 | 67 | [25] |
| 2023 | 5 × 1000 | 29th GaSb | 1278 | cleaved | 10 | 30 | 56 | [10] |
| 2024 | 2.4 × 3000 | 1st order Amorphous Silicon | 3639 | 0.07%/99% | 47.02 | 46.7 | 6.72 | This work |

## Conclusion

A novel approach for high-performance GaSb-based laterally coupled distributed feedback (LC-DFB) lasers, utilizing an α-Si dielectric Bragg grating and a precision trapezoidal waveguide etching technique, is used to reinforce the effectiveness of regrowth-free fabrication process that is particularly advantageous for materials systems like GaSb that lack a suitable etch stop layer. The resulting LC-DFB lasers exhibit superior high power and spectral purity at a wavelength of 1.95 μm, as detailed in **Table 1**. The high output power is attributed to several factors: the reduced optical loss and limited lateral current spreading credited to the insulating dielectric grating, together with the 3 mm long cavity design which allows for more amplification of light and thereby increases the gain, and the anti-reflection/high-reflection (AR/HR) coating design, which optimizes light field distribution and maximizes power extraction efficiency. Additionally, the long cavity design and the uniform light field contribute to the linewidth reduction[26]. The exceptional single-mode performance is a result of the high refractive index contrast of the dielectric grating, which facilitates strong longitudinal mode selectivity, and the stable coupling coefficient (κ) provided by the deterministic nature of the dielectric grating structure. This stable κ ensures consistent and predictable device performance. The regrowth-free fabrication scheme simplifies the manufacturing process, reduces production costs, and enhances the versatility of semiconductor laser manufacturing. By eliminating the need for an etch stop layer, this approach broadens the range of gain materials in which high-performance LC-DFB lasers can be fabricated.